\documentclass[aps,prl,twocolumn,superscriptaddress,nopacs,citeautoscript]{revtex4}

\usepackage{graphicx}
\usepackage{amsfonts}
\usepackage{amssymb}
\usepackage{color}

\newcommand\ea{\textit{et al.}}

\begin{document}


\title{Reply to ``Comment on arXiv:1012.1484v1 Structural origin of apparent Fermi surface pockets in angle-resolved photoemission of Bi$_2$Sr$_{2-x}$La$_x$CuO$_{6+\delta}$ by King {\it et al.} ''}


\author{P.~D.~C.~King}
\affiliation{School of Physics and Astronomy, University of St Andrews, North Haugh, St Andrews, KY16 9SS, United Kingdom}

\author{J.~A.~Rosen}
\affiliation{Department of Physics and Astronomy, University of British Columbia, Vancouver, British Columbia V6T 1Z1, Canada}

\author{W.~Meevasana}
\affiliation{School of Physics and Astronomy, University of St Andrews, North Haugh, St Andrews, KY16 9SS, United Kingdom}
\affiliation{School of Physics, Suranaree University of Technology, Nakhon Ratchasima, 30000 Thailand}

\author{A.~Tamai}
\author{E.~Rozbicki}
\affiliation{School of Physics and Astronomy, University of St Andrews, North Haugh, St Andrews, KY16 9SS, United Kingdom}

\author{R.~Comin}
\author{G.~Levy}
\author{D.~Fournier}
\affiliation{Department of Physics and Astronomy, University of British Columbia, Vancouver, British Columbia V6T 1Z1, Canada}

\author{Y.~Yoshida}
\author{H.~Eisaki}
\affiliation{National Institute of Advanced Industrial Science and Technology, Tsukuba, Ibaraki 305-8568, Japan}

\author{K.~M.~Shen}
\affiliation{Laboratory of Atomic and Solid State Physics, Cornell University, Ithaca, New York 14853, USA}

\author{N.~J.~C.~Ingle}
\affiliation{AMPEL, University of British Columbia, Vancouver, British Columbia V6T 1Z1, Canada}
\author{A.~Damascelli}
\affiliation{Department of Physics and Astronomy, University of British Columbia, Vancouver, British Columbia V6T 1Z1, Canada}
\affiliation{Quantum Matter Institute, University of British Columbia, Vancouver, British
Columbia V6T 1Z4, Canada}

\author{F.~Baumberger}
\email{fb40@st-andrews.ac.uk}
\affiliation{School of Physics and Astronomy, University of St Andrews, North Haugh, St Andrews, KY16 9SS, United Kingdom}

\maketitle

In our recent letter~\cite{King:arXiv:1012.1484:(2010)}, we combined polarization-dependent angle-resolved photoemission spectroscopy (ARPES) and low-energy electron diffraction (LEED) measurements to show that the apparent closed Fermi surface pockets recently reported by ARPES from under-doped (UD) Bi$_2$Sr$_{2-x}$La$_x$CuO$_{6+\delta}$ (La-Bi2201)~\cite{Meng:Nature:462(2009)335--338} arise from structural artifacts characteristic to these compounds, and do not represent the generic Fermi surface topology of UD cuprates. In a comment on this work, Zhou~\ea~\cite{Zhou:arXiv:1012.3602:(2010)} argue against this conclusion. Their concerns can be summarized by two points: (1) taking a second superstructure vector $\mathbf{q}=(q_2,q_2)\frac{\pi}{a}$ with $q_2=0.092$ (Fig.~1 of Ref.~\cite{Zhou:arXiv:1012.3602:(2010)}), neither the number of bands nor the topology of the Fermi surface pockets can be reproduced; and (2) the relative intensity of the different diffraction replica (DR) does not simply scale with their diffraction order. We address these points in turn below.

(1) Taking a second superstructure vector of $q_2=0.092$ is not representative of the structural modulations present in the UD La-Bi2201 compounds investigated by us or by Zhou and colleagues~\cite{Meng:Nature:462(2009)335--338}. As clearly shown from our LEED measurements (Fig.~4a of Ref.~\cite{King:arXiv:1012.1484:(2010)}), the second superstructure is $q_2=0.12$. This is, in fact, commensurate with the $\mathbf{q}_1$ superstructure in these compounds, which simplifies the situation considerably from the picture put forward in Fig.~1c of Zhou~\ea~\cite{Zhou:arXiv:1012.3602:(2010)}.  

We stress that using a tight-binding model fitted to only the main band, and taking superstructure vectors determined {\it independently} from our LEED, without any adjustment, we get good agreement over an extended {\bf k}-space range with the experimental Fermi surfaces measured both by us and by Zhou and colleagues~\cite{Meng:Nature:462(2009)335--338} for UD and optimally-doped (OP) La-Bi2201, as well as OP (Pb,La)-Bi2201, as is evident in Figs.~1, 2, and 4 of Ref.~\cite{King:arXiv:1012.1484:(2010)}. In particular, {\it all} pocket-like features in the UD La-Bi2201 and OP (Pb,La)-Bi2201 are simply explained by this model, as is their absence in OP La-Bi2201. The slight deviations from our model for the `LPS' pocket pointed out by Zhou~\ea~\cite{Zhou:arXiv:1012.3602:(2010)} are consistent with residual spectral weight being contributed by a nearby DR of the main band. We note that this provides a natural explanation for the small changes in shape and size between the different pockets observed experimentally in Ref.~\cite{Meng:Nature:462(2009)335--338}, which itself was an inconsistent feature of the model of intrinsic pockets proposed there.

(2) The relative intensity of DR is strongly photon energy dependent and does not simply scale with their order, as commonly observed in ARPES measurements of DR in the single-, double-, and triple-layer Bi-based cuprates~\cite{Damascelli:Rev.Mod.Phys.:75(2003)473--541,Mans:Phys.Rev.Lett.:96(2006)107007,Feng:Phys.Rev.Lett.:86(2001)5550--5553,Borisenko:Nature:431(2004)--,Feng:Phys.Rev.Lett.:88(2002)107001,Ideta:Phys.Rev.Lett.:104(2010)227001}. Such variations are also apparent in our ARPES spectra of (Pb,La)-Bi2201 (Fig.~2 of Ref.~\cite{King:arXiv:1012.1484:(2010)}), for which Zhou~\ea~\cite{Zhou:arXiv:1012.3602:(2010)} agree that the multiple bands and apparent Fermi pockets result from two co-existing superstructures. In these measurements, most first-order DR are visible, as are a few second-order ones, just as for our measurements from both UD and OP La-Bi2201. Matrix element variations can even cause a DR to be more intense than the original band. This is evident in previous studies of Bi2201~\cite{Mans:Phys.Rev.Lett.:96(2006)107007}, including Zhou~\ea's own data: for example, $I_{MB-\mathbf{q}_1}>I_{MB}>I_{MB+\mathbf{q}_1}$ in Fig.~2b of Ref.~\cite{Meng:Nature:462(2009)335--338}, where $I_{MB(\pm\mathbf{q}_1)}$ denotes the intensity of the main band or its $\pm\mathbf{q}_1$ DR. It is therefore inconsistent to argue that the shadow band itself must be clearly visible in order for its DR to exist, as claimed by Zhou~\ea~\cite{Zhou:arXiv:1012.3602:(2010)}. Indeed, in their own study~\cite{Meng:Nature:462(2009)335--338}, the clearly-observed `LS' band is attributed to the $-\mathbf{q}_1$ DR of the shadow band, while the shadow band itself cannot be discerned. 

Given that our LEED measurements unambiguously show the presence of second superstructure vectors in the underdoped La-Bi2201 samples, diffraction replica corresponding to these superstructures must be expected. We can identify the majority of first-order and some second-order replica bands corresponding to the multiple superstructures characteristic of the investigated compounds. While the relative intensity of the bands does show a complex variation due to matrix element effects, the precise superstructure vectors determined {\it independently} from LEED explain the {\it entire} Fermi surface topology at all doping levels measured in ARPES, both by us and by Zhou and colleagues~\cite{Meng:Nature:462(2009)335--338}, including all of the apparent Fermi surface pockets. In our opinion, this makes it very difficult to argue that the observed pockets are intrinsic to the electronic structure of a doped CuO$_2$ plane, as claimed in Ref.~\cite{Meng:Nature:462(2009)335--338}.

\bibliographystyle{apsrev}

\end{document}